\documentclass[]{aa} 

\usepackage{graphicx}
\usepackage{natbib}
\usepackage{ulem}
\usepackage{subfig}
\usepackage{float}
\usepackage[usenames,dvipsnames]{xcolor}
\usepackage{multirow}
\usepackage{footnote}
\usepackage[bottom]{footmisc}
\usepackage{ulem}
\usepackage{xltabular}
\usepackage{longtable}
\usepackage{lipsum} 
\usepackage{pdflscape}
\usepackage{hyperref}
\usepackage{xcolor}
\hypersetup{
        colorlinks   = true, 
        urlcolor     = black, 
        linkcolor    = blue, 
        citecolor   = blue 
}
\usepackage[all]{hypcap}
\usepackage{txfonts}
\newcommand{\teff}{T_\mathrm{eff}}
\newcommand{\logg}{\mathrm{log}\,g}

\begin{document}

  \title{Exploring the stellar properties of M dwarfs with high-resolution spectroscopy from the optical to the near-infrared \thanks{Table 2 is only available in electronic form
at the CDS via anonymous ftp to cdsarc.u-strasbg.fr (130.79.128.5) or via http://cdsweb.u-strasbg.fr/cgi-bin/qcat?J/A+A/}}

\titlerunning{High-resolution spectroscopy of M dwarfs}
 \authorrunning{Rajpurohit et al.}

\author{A. S. Rajpurohit \inst{1},  F. Allard \inst{2},  S. Rajpurohit\inst{3}, R. Sharma\inst{1}, G. D. C. Teixeira \inst{4, 5}, O. Mousis \inst{6}, R. Kamlesh\inst{7}}

\institute{Astronomy \& Astrophysics Division, Physical Research Laboratory, Ahmedabad 380009, India
\email{arvindr@prl.res.in}
\and
Univ Lyon, Ens de Lyon, Univ Lyon1, CNRS, Centre de Recherche Astrophysique de Lyon UMR5574, F-69007, Lyon, France
\and
Institut f\"{u}r Theoretische Physik, Universit\"{a}t G\"{o}ttingen, 37077 G\"{o}ttingen, Germany
\and
Instituto de Astrof\'{i}sica e Ci\^{e}ncias do Espa\c{c}o, Universidade do Porto, CAUP, Rua das Estrelas, 4150-762 Porto, Portugal
\and
Departamento de F\'{i}sica e Astronomia, Faculdade de Ci\^{e}ncias, Universidade do Porto, Rua Campo Alegre, 4169-007 Porto, Portugal
\and
Aix Marseille Universit\'{e}, CNRS, LAM (Laboratoire d'Astrophysique de Marseille) UMR 7326, 13388, Marseille, France
\and
Dipartimento di Fisica e Astronomia, Via Gobetti 93/2, 40131, Bologna, Italy
}

\date{\today}
 
  \abstract
   {Being the most numerous and oldest stars in the galaxy, M dwarfs are objects of great interest for exoplanet searches. The presence of molecules in their atmosphere complicates our understanding of their atmospheric properties. But great advances have recently been made in the modeling of M dwarfs due to the revision of solar abundances.}
   {We aim to determine stellar parameters of M dwarfs using high resolution spectra (R$\sim$90 000) simultaneously in the visible and the near-infrared. The high resolution spectra and broad wavelength coverage provide an unique opportunity to understand the onset of dust and cloud formation at cool temperatures. Furthermore, this study will help in understanding the physical processes which occur in a cool atmospheres, particularly, the redistribution of energy from the optical to the near-infrared.}
   {The stellar parameters of M dwarfs in our sample have been determined by comparing the high resolution spectra both in the optical and in the near-infrared simultaneously observed by CARMENES with the synthetic spectra obtained from the BT-Settl model atmosphere. The detailed spectral synthesis of these observed spectra both in the optical and in the near-infrared helps to understand the missing continuum opacity.}
   {For the first time, we derive fundamental stellar parameters of M dwarfs using the high resolution optical and near-infrared spectra simultaneously. We determine $\teff$, $\logg$ and [M/H] for 292 M dwarfs of spectral type M0 to M9, where the formation of dust and clouds are important. The derived $\teff$ for the sample ranges from 2300 to 4000 K, values of $\logg$ ranges from 4.5 $\le$ $\logg$ $\le$ 5.5 and the resulting metallicity ranges from  -0.5 $\le$ [M/H] $\le$ +0.5. We have also explored the possible differences in $\teff$, $\logg$ and [M/H] by comparing them with other studies of the same sample of M dwarfs.}
   {}


   \maketitle
%

\section{Introduction}
The large number of M dwarfs present in our Galaxy (70$\%$ of the Galactic stellar population \citep{Bochanski2010}) makes them one of the most important stellar populations. The low mass range of M dwarfs (0.6 $M_{\odot}$ to the hydrogen burning limit of about 0.075 $M_{\odot}$ ) and the small size make them suitable candidates to detect planets around them in the habitable zone. Recent studies show that M dwarfs also host brown dwarfs and exoplanets  \citep{Bonfils2012,Anglada2016,Gillon2017} which makes them important targets to study and understand exoplanet formation.

Unlike other stellar objects, the stellar properties of M dwarfs varying a lot from early to late M dwarfs, that is, from M0 to M9. Though there are huge numbers of M dwarfs in our Galaxy, we still lack a homogeneous sample with respect to their particular age and metallicity. Because of their intrinsic faintness it is difficult to obtain good signal-to-noise ratio (S/N) and high resolution spectra both in the visible and in the near-infrared (NIR) simultaneously. Spectrographs such as HARPS \citep{Mayor2003} or HARPS-N \citep{Cosentino2012} provide high resolution optical spectra whereas the CRIRES spectrograph \citep{Kaeufl2004} provides high resolution NIR spectra of M dwarf stars. Results from such spectrographs have given us hints on the differences that exist between the NIR and the optical spectra of cool M dwarfs and about  different features which could be used to characterize the whole sequence of M dwarf stars. Recently, CARMENES  \citep{Quirrenbach2014} started providing simultaneous high resolution (R $\sim$ 90, 000) observation of M dwarfs both in the visible (0.52 to 0.96 $\mu$m) and in  the near-infrared (0.96 to 1.71 $\mu$m) wavelengths. Future high resolution spectrographs such as SPIRou \citep{Cersullo2017} and HPF \citep{Mahadevan2012} will further provide good quality, high signal-to-noise ratio spectra of M dwarfs. 

The temperatures of M dwarf atmosphere are cool enough to form diatomic and triatomic molecules. The presence of these molecules, such as SiH, CaH, TiO, VO, CrH, MgH, OH, CO, CaOH, H$_2$O, and FeH, can be seen both in the optical and in the near-infrared (NIR) spectra of M dwarfs. \cite{Tsuji1996a} identified dust formation by recognising the condensation temperatures of hot dust grains occurring in the line-forming  layers of M-dwarf atmospheres. In particular, the temperature of the outermost layers in M dwarfs, with spectral type M5 or later, is cool enough to form dust and clouds. This causes the weakening or vanishing  of TiO and VO molecular bands from the optical spectra of late M dwarfs. Thus, the continuum formed by the atoms in M dwarfs, is much weaker than the molecular or dust background, contrary to the hotter stars. Thus, in M dwarfs, the molecular pseudo-continuum which is made of millions transitions dominates the atomic classical continuum by the orders of magnitudes \citep{Allard1990,Allard1995}. These complex molecules and dust grains in M-dwarf atmospheres make the access to the atomic continuum nearly impossible, increasing the difficulty of the atmosphere modeling and thus, the difficulty of determining the stellar atmospheric parameters.    
   
M dwarfs are the prime targets for finding the planets in the habitable zone. The properties of these planets directly depend on the properties of host stars \citep{Santos2004,Mann2013c,Gaidos2014,Souto2017}. Thus, it is crucial to determine stellar parameters more precisely and accurately. It is very important to determine the stellar parameter of M dwarfs in the optical and in the NIR simultaneously to overcome any discrepancy and biases,  resulting from their complex atmospheres. 

Various approaches have been made by different groups using different methods, but till today, stellar parameters such as effective temperature ($\teff$), surface gravity ($\logg$), and metallicity ([M/H]) are not yet well determined for M dwarfs with great accuracy. These stellar parameters are still model dependent to some extent for M dwarfs. In the past, the $\teff$ of M dwarfs were determined using broadband photometry and blackbody approximations due to the lack of reliable models. Those estimates of the $\teff$ of M dwarfs are not reliable, given the complex and broad molecular absorptions in their atmospheres. Also, attempts have been made to determine the $\teff$ for nearby K and M dwarfs, based on interferometrically determined radii and bolometric fluxes from photometry \citep{Boyajian2012}. As interferometric measurements are not currently possible for the cool and fainter M dwarfs, \cite{Boyajian2012} restricted their work to early M dwarfs, in other words, up to the spectral type M5. 

Atmosphere modeling of the cool low-mass stars and the substellar objects has developed \citep{Allard2012,Allard2013} in recent  decades due to the parallel improvement of computing capacities. More realistic model atmospheres and synthetic spectra for the very low mass stars (VLMs), brown dwarfs and extrasolar planets, have been made possible. Models, such as, BT-Settl \citep{Allard2013}  have succeeded in modeling various complex molecular absorption bands by incorporating the revised solar abundances along with updated atomic and molecular line opacities which govern the spectral energy distribution (SED) of M dwarfs. These new updated models, now include the dust and cloud formation \citep{Allard2013} which is important for the cool M-dwarfs and thus, yields promising results, which explain the stellar-to-substellar transition. Recently, \cite{Rojas2010, Roja2012} have used the BT-Settl model to determine temperatures and metallicities by measuring the equivalent widths (EW) of  Na I, Ca I, and the H$_2$O-K2 index. \cite{Mann2015} determined the $\teff$ of the M dwarfs following the approach of \cite{mann2014}, using the optical spectra and the BT-Settl model. Moreover, \cite{Souto2017} used the MARCS model \citep{Gustafsson2008} and performed the detailed NIR chemical abundance analysis, observed by SDSS-IV-Apache Point Observatory Galactic Evolution Experiment \cite[APOGEE,][]{Majewski2015} for early M dwarfs. The adopted $\teff$ in \cite{Souto2017} is based on the photometric calibrations for M dwarfs by \cite{Mann2015} for the V-J and R-J colors. \cite{Rajpurohit2013,Rajpurohit2014,Rajpurohit2018} have used both low and medium resolution spectra to determine  the $\teff$, log g and [M/H] of the M dwarfs, M subdwarfs, using the most recent BT-Settl model atmosphere. Recently, \cite {Passegger2018} used PHOENIX-ACE model atmospheres to estimate the fundamental stellar parameters of M dwarfs by comparing them to the high resolution optical spectra of M dwarfs. 

In this paper, the high-resolution visible and NIR spectra of 292 M dwarfs were obtained from CARMENES, to determine their atmospheric parameters ($\teff$, ${\logg}$ and [M/H]) simultaneously by using the most recent BT-Settl model grid. For the first time the BT-Settl models are used to compare simultaneous high-resolution visible and NIR spectra of M dwarfs. We briefly describe the sample selection in Section \ref{obs}. Description of the BT-Settl model atmosphere is given in Section \ref{models}. Results are presented in Section \ref{parameters}. Analysis and discussion are presented in Section \ref{analysis}, which describe the behavior of the models, while comparing with observations and their stellar parameters. Summary of the paper is presented in Section \ref{Summary}.

\begin{figure*}[!htp]
\centering
\vspace{-3.0cm}
\subfloat{\includegraphics[height=18.1cm, width=15cm,angle=-90]{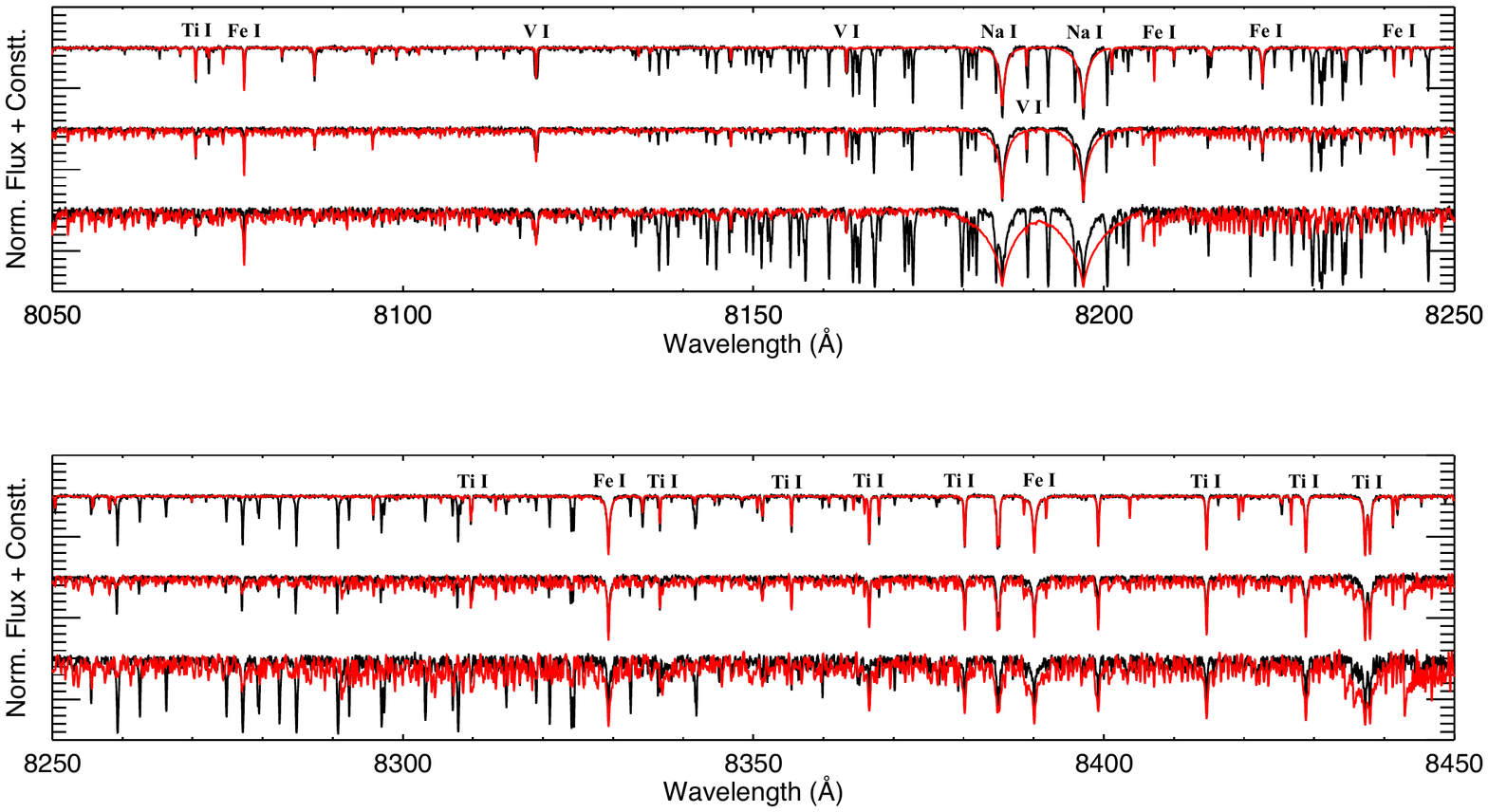}}
\vspace{-6.0cm}
\qquad
\subfloat{\includegraphics[height=18.1cm, width=15cm,angle=-90]{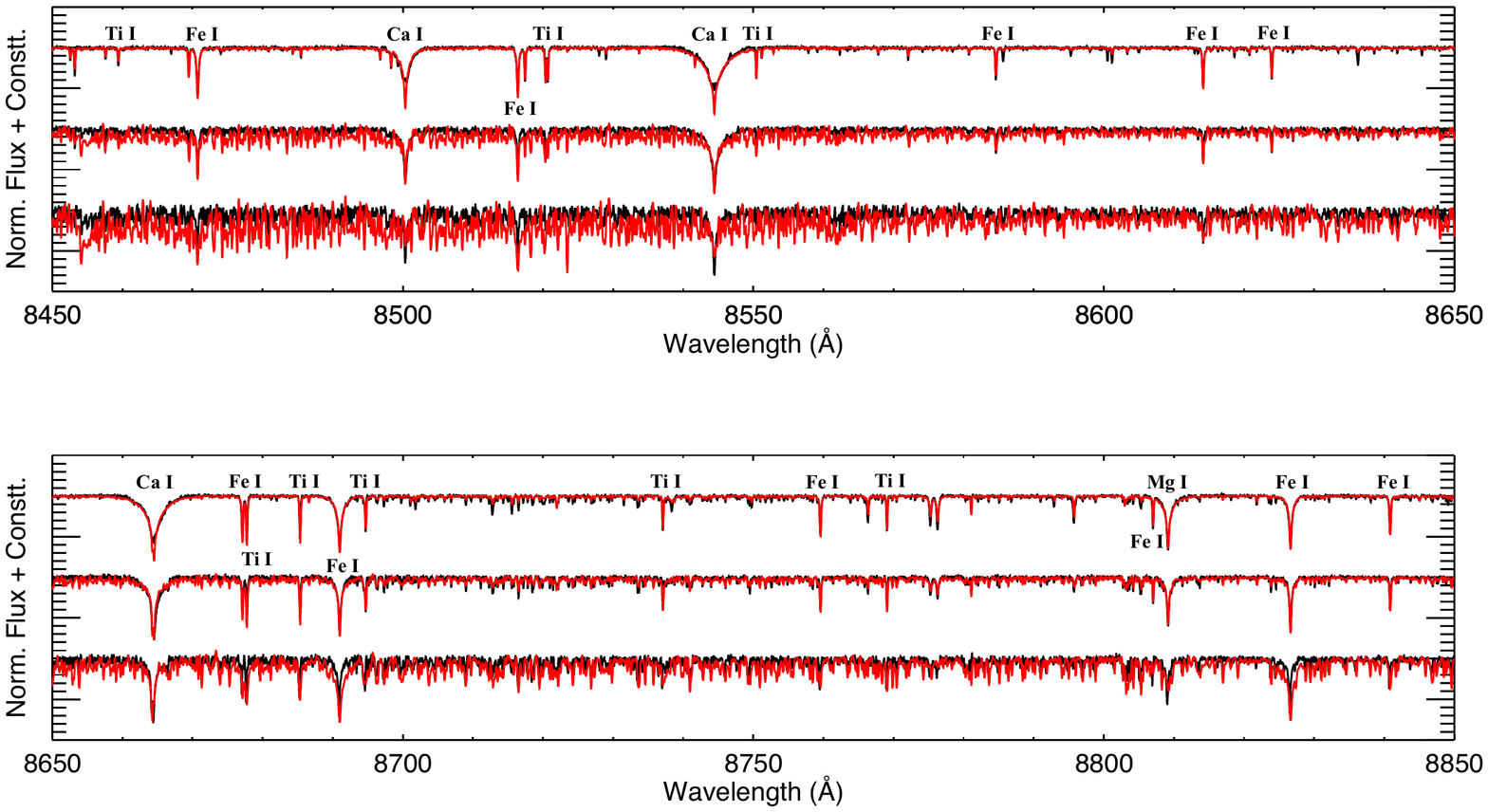}}
\vspace{-3.0cm}
\caption{CARMENES spectra of GJ 180 (M1.0, Top), Ross 128 (M4.0, middle), HD 180617 (M8.0, bottom) in black is compared with the best-fit BT-Settl model (red). The main spectral features which includes atomic lines such as Fe I, Ca I, Na I, K I, Si I, Mg I, Al II, along with some hydride bands such as those of FeH and OH can be seen. We used mainly \cite{Tinney1998,Reiners2017} for the spectral features recognition and labeling. Their best value of  $\teff$, [M/H] and $\logg$ is given in Table \ref{Table2}.}
\label{Fig1}
\end{figure*}

\begin{figure*}[!htp]
\centering
\vspace{-3.0cm}
\subfloat{\includegraphics[height=18.1cm, width=15cm,angle=-90]{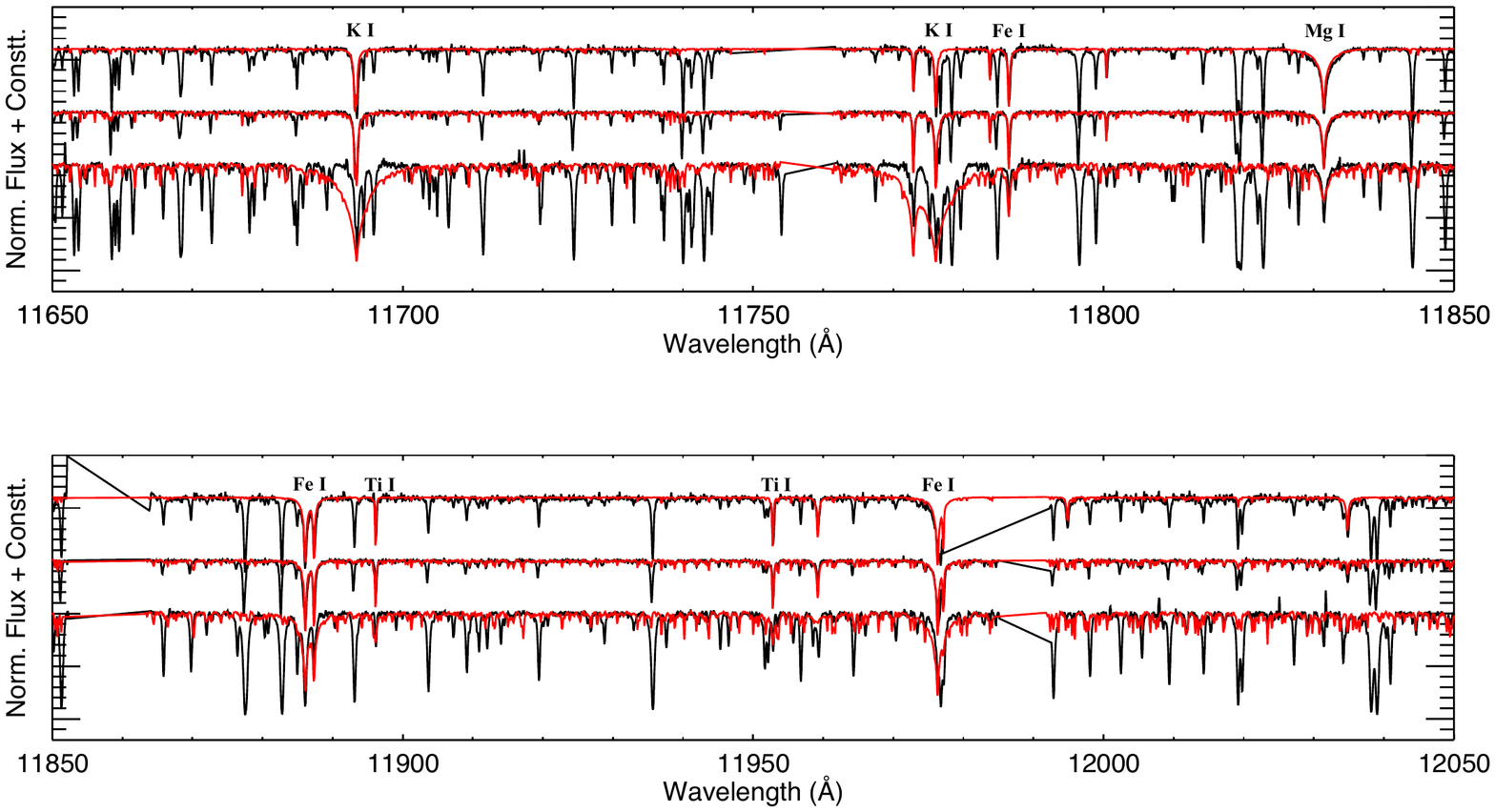}}
\vspace{-6.0cm}
\qquad
\subfloat{\includegraphics[height=18.1cm, width=15cm,angle=-90]{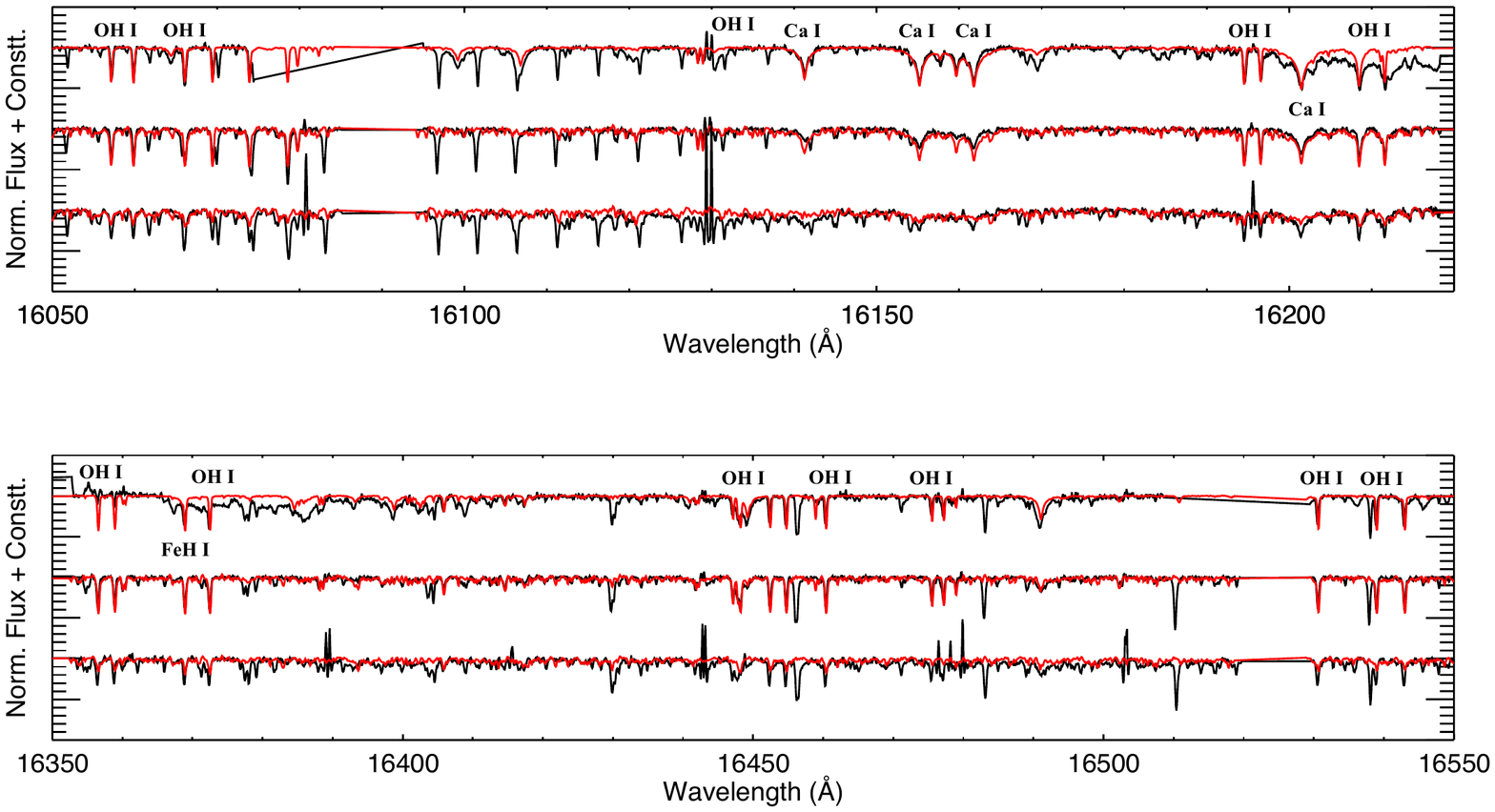}}
\vspace{-3.0cm}
\caption{Same as Fig~\ref{Fig1} but in NIR regime.}
\label{Fig2}
\end{figure*}

\section{Sample selection}
\label{obs}
We have obtained 292 M dwarfs CARMENES spectra  through their GTO agreement, as mentioned in \cite{Reiners2017}. Regarding the details of CARMENES targets, the reader is referred to \cite{Reiners2017}. The details of CARMENES data reduction are described in \cite{Caballero2016b} and \cite{Nidever2015}. The M dwarfs analyzed in this study, were selected based on their late spectral type and brightness. The typical limiting magnitude of CARMENES survey for the J-band is J = 12 mag. The heliocentric distances ranging from 1.8 pc to 39.1 pc, and the proper motions from 0.03 arcsec a$^-$$^1$ to 10.6 arcsec a$^-$$^1$ \citep{Passegger2018}. As suggested by \cite{cortes2017} and \cite{Passegger2018}, the typical age of the stars in our sample is 5 Gyr. The spectral type and the photometry were compiled using the Simbad and Vizier catalog, access through Centre de Donnees astronomiques de Strasbourg. The spectral type of M dwarfs used in our study, range from M0.0 to M9.5. These spectra have not been corrected for terrestrial absorptions lines.

The SED of M dwarfs is mainly governed by diatomic and triatomic molecules. The optical spectra (0.64 to 0.96 $\mu$m) of M dwarfs are mainly dominated by the diatomic metal oxides, such as, titanium oxide (TiO) and vanadium oxide (VO). The strength of TiO bands decrease as we go from early to late M dwarfs, whereas, VO produce more diffuse absorption toward the redder wavelengths. The most dominant TiO "$\gamma$"  bandheads in the optical spectra of M dwarfs are around 8205.8, 8250.6, 8289.0, 8302.9, 8334.5, 8375.5, 8386.5, 8419.5, 8442.3, 8451.7, 8457.1, 8471.6, 8505.5, 8513.1, 8558.4, 8569.4, 8859.6, 8868.5, 8937.4, 8949.8, 9014.6, and 9094.5 $\AA$. The VO bandheads in M dwarfs are located at  7896.0, 7899.6, 7918.4, 7928.5, 7938.9, 7947.7, 7960.1, 7967.2, 7973.1, 7982.1, 8520.9, 8537.7, 8572.8, 8575.3, 8590.7, 8597.2, 8604.0, 8624.0, 8648.6, 8657.9, and 8666.6  $\AA$ \citep{Tinney1998}. These molecular absorption bands can be seen over the entire spectral sequence of M dwarfs. These metal oxides along with some hydrides, such as CaH are the major source of the opacity in the M dwarfs. As we go from early to late M dwarfs, the strengthening of hydride bands and collision-induced absorption (CIA) by H$_2$, and the broadening of atomic lines \citep{Allard1997} occurs. The NIR spectra  (0.96 to 1.71 $\mu$m), is the region where the presence of broad and complex molecular absorption makes it difficult to identify various weak atomic absorption features in the spectra. H$_2$O, FeH, CO and OH are the dominant spectral features in the NIR spectra of M dwarfs along with neutral metals. As one goes from early to late M dwarfs, H$_2$O and CO become stronger with decreasing $\teff$. Molecular species, such as FeH and OH, produce more diffuse absorption in the NIR as compared to TiO and VO in the optical. 

All the observed M dwarfs spectra show strong alkali lines, both in the visible and in the NIR wavelength range (0.64 to 1.71 $\mu$m). As expected from their high surface gravity, these atomic lines are massively pressure broadened. For example, Figs.~\ref{Fig1} and~\ref{Fig2} show the dominant atomic features, such as, Fe I, Ti I, Al I, Ca I, Na I, K I, and Mg I, which are visible throughout the M dwarfs spectral sequence (for more details of spectral features, see Table \ref{table1}). All these atomic lines are prominent in almost all of the spectra. However, it is difficult to measure the intensities of these lines in the region where strong atmospheric and molecular absorption are present. The atomic lines, such as, Ca I, K I, Na I and Mg I are relatively free from any blends and uncontaminated by the telluric lines. These are the ideal features to study their dependency on various stellar parameters of M dwarfs. These lines get broadened as one goes from early to later M dwarfs and their typical equivalent widths are of several Angstroms. The strength of these atomic features depends on various stellar parameters, such as, $\teff$, [M/H] and luminosity. We refer the reader to  \cite{Tinney1998, Rajpurohit2014}
and \cite {Reiners2017} for more details of the spectral features and their labeling over the entire observed wavelength region of CARMENES spectra.

\section{BT-Settl model}
\label{models}
The BT-Settl models \citep{Allard2012,Allard2013} used in this study, were calculated using the "PHOENIX" radiative transfer code \citep{Hauschildt1997,Allard2001}. These model are based on assumptions, such as the convection using the mixing-length theory, the hydrostatic and the chemical equilibrium, and a line by line treatment of the opacities. \cite{Allard2012} have included the dust formation as well as gravitational settling \citep{Allard2003,Allard2012}. These model also takes into account a number of microphysical process as well as gravitational settling \citep{Allard2003,Allard2012}. The CO line list used in the model is taken from \cite{Goorvitch1994a,Goorvitch1994b} whereas the updated water vapor line list is taken from Barber and Tennyson (BT2) \citep{Barber2006}. Detailed profiles of the alkali lines are also used \cite{Allard2007} in the current version of the BT-Settl model. The TiO, VO and CaH line list used in the BT-Settl model are from \cite{Plez1998,Plez2008}, MgH by \cite{Skory2003}, and \cite{Weck2003}, H$_2$ Collision Induced Absorption (CIA) by \cite{Borysow2001} and \cite{Abel2011}, CO$_2$ by \cite{Tashkun2004}, FeH and CrH by \cite{Chowdhury2006} and \cite{Dulick2003}, NH$_3$ by \cite{Yurchenko2011}. In BT-Settl model atmosphere, the dust is assumed to be formed when the supersaturation ratio S $\ge$ 1.00. For each layer in the photosphere, the dust grains mean size and number densities were calculated by comparing the timescales for the condensation, coagulation, and gravitational settling  with the time scale for mixing due to convective overshooting. For the opacity treatment, 55 condensates were included, such as ZrO$_2$, Al$_2$O$_3$, CaTiO$_3$, Ca$_2$Al$_2$SiO7, MgAl$_2$O4, Ti$_2$O$_3$, Ti4O7, Ca$_2$MgSi$_2$O7, CaMgSi$_2$O6, CaSiO$_3$, Fe, Mg$_2$SiO4, MgSiO$_3$, Ca$_2$SiO4, MgTiO3, MgTi$_2$O5, Al$_2$Si$_2$O13, VO, V$_2$O3, and Ni, to name a few \citep{Allard2014}. We have assumed that the dust grains are spherical, homogeneous and distributed according to a log-normal distribution. The grid of the BT-Settl model extend $\teff$ from 300 to 7000 K, in steps of 100 K, from log g = 2.5 to 5.5 dex, in steps of 0.5 dex, and with [M/H] = -2.5 to +0.5 dex, in steps of 0.5 dex. BT-Settl models also includes the alpha-enhancement and the latest solar abundances by \cite{Caffau2009,Caffau2011}.

\begin{table}[ht!]
\caption{Wavelength regions and lines used for ${\chi}^2$ calculations}
\centering
\begin{tabular}{ | c | p{5cm} |}\hline
Line/band &Wavelength ({\AA}) \\

\hline
\multirow{ 2}{*}{Ti I}     & $8611.91,8684.23,8692.32,8734.71,$\\
 &$     8766.68,8993.64, 9029.86, 9708.28,$\\
&$   9721.55, 9731.02, 9834.62,10037.3,$\\
&$   10399.6,10498.9,10587.5,   10610.6,$\\
&$   10664.5,10679.9, 10735.9,  10735.8,$\\
&$   10777.9, 11896.3,11976.8,  12825.4,$\\
&$      12834.9,15548.0,15607.0,15703.0,$\\
&$      15719.8,16639.6$\\
 
 \hline
 
\multirow{ 3}{*}{Fe I}    
&$      7513.22,        7585.94,
        7750.45,        7782.77,$\\
&$      7834.26,        7915.11,
        7939.27,        7948.10,$\\
&$      7948.10,        8001.26,
        8048.26,        8077.33,$\\
&$      8222.64,  8241.39,
        8329.24,        8390.09,$\\
&$      8470.77,        8516.42,        
        8584.58,        8614.20,$\\
&$      8623.95,   8664.63,
        8677.17,        8691.05,$\\
&$      8759.63, 8807.04,
        8826.61, 8840.88,$\\
&$      8979.23, 9003.04,
        9121.78,        9148.59,$\\
&$      10398.7,        11124.3,
        11425.7,        11489.2,$\\
&$      11597.9,        11613.0,
        11641.6,        11693.4,$\\
&$      11786.6,        11976.5,
        12883.3,        15298.8,$\\
&$      16490.9$\\

\hline  Al I     & $13155.0$ \\
\hline
\multirow{ 2}{*}{Ca II}  & $7204.26,7328.26,7891.18,8500.39,$\\
                                  & $8544.42,8664.63,10346.7,12819.5,$\\
                                  & $16141.3,16155.0,16201.3$\\
\hline
\multirow{ 2}{*}{K I}     & $7668.38,7701.13,11693.0.$\\
                                  & $11772.9,12435.7,12525.6,15168.5$\\
                                   &$15172.4$   \\
\hline
\multirow{ 2}{*}{Na I}     & $8183.3,8194.8,11385.0,$\\
                                        &$11410.0$ \\
\hline
\multirow{ 2}{*}{Mg I}     & $8809.2,11831.4,14890,15753.2,$\\
                                    & $15770.0$ \\
\hline 
\multirow{ 2}{*}{OH}
&       $15134.9, 15135.4,
        15149.8, 15149.7, $\\
&       $15268.9,       15149.7,
        15539.7, 15561.9,$\\
&        $15572.9,16208.6,
        16211.6, 16369.1,$\\
&       $16460.5,       16530.6,
        16543.1,16586.3,$\\
&       $16883.4,       16888.8,
        16899.6,        17054.6,$\\
&$      17070.5,        17086.5$\\

\hline
\end{tabular}
\label{table1}
\end{table}

\section{Results}
\label{parameters}
We adapted the method, described in \cite{Rajpurohit2012a,Rajpurohit2018} to determine the stellar parameters of the M dwarfs sample used in this study. We have used the most recent atmosphere models and spectroscopic informations covering both in the optical and in the NIR ranges to derive their stellar parameters. As a first step, we normalized both observed and synthetic spectra, by applying a boxcar filter to remove the absorption features, by dividing each spectrum into many short wavelength intervals. This was done by fitting the polynomial of the second order to the data. We then apply a multiple iterative process, until a good continuum fit for the whole spectrum is obtained. Secondly, we degraded the resolution of each synthetic spectra with a Gaussian profile with the measured instrumental resolution at the observed resolution and we then rebin the outcome at each wavelength point of the observed spectra. For the first estimate, we performed a $\chi^2$ minimisation test using the set of model atmosphere grid covering the range of 2200 K $\ge$ $\teff$ $\ge$ 4000 K in a step of 100 K, 4.5 $\ge$ $\logg$ $\ge$ 5.5, and [M/H] = -0.5 to +0.5 in a step of 0.5 dex. This procedure includes the calculation of the difference between the flux of the observed and the synthetic spectra at every wavelength point. Thereby, we obtained the sum of the squares of these differences for each model in the grid, and finally selects the best fit model for each source. We retained the best-match values of $\teff$, $\logg$ and [M/H] as our first guess values on these parameters. All the three parameters ($\teff$, $\logg$ and [M/H]) have been kept free during this step, to remove any biases in the parameter space. The wavelength regions along with atomic and molecular features, used for the $\chi^2$ calculation, are given in the Table \ref{table1}.

\begin{figure*}[!htp]
\centering
\vspace{-2.0cm}
\subfloat{\includegraphics[height=8.8cm, width=9.3cm]{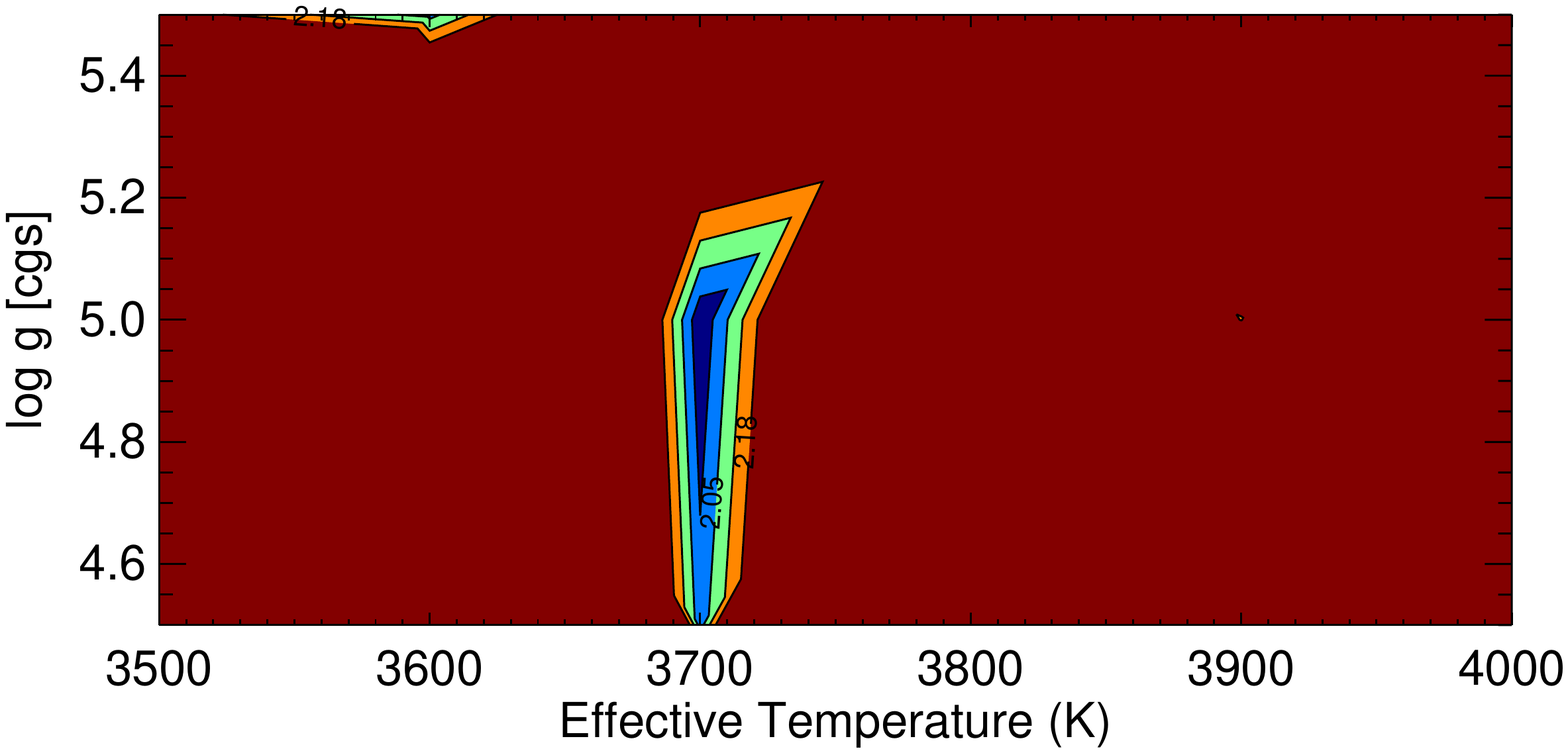}}
\subfloat{\includegraphics[height=8.8cm, width=9.3cm]{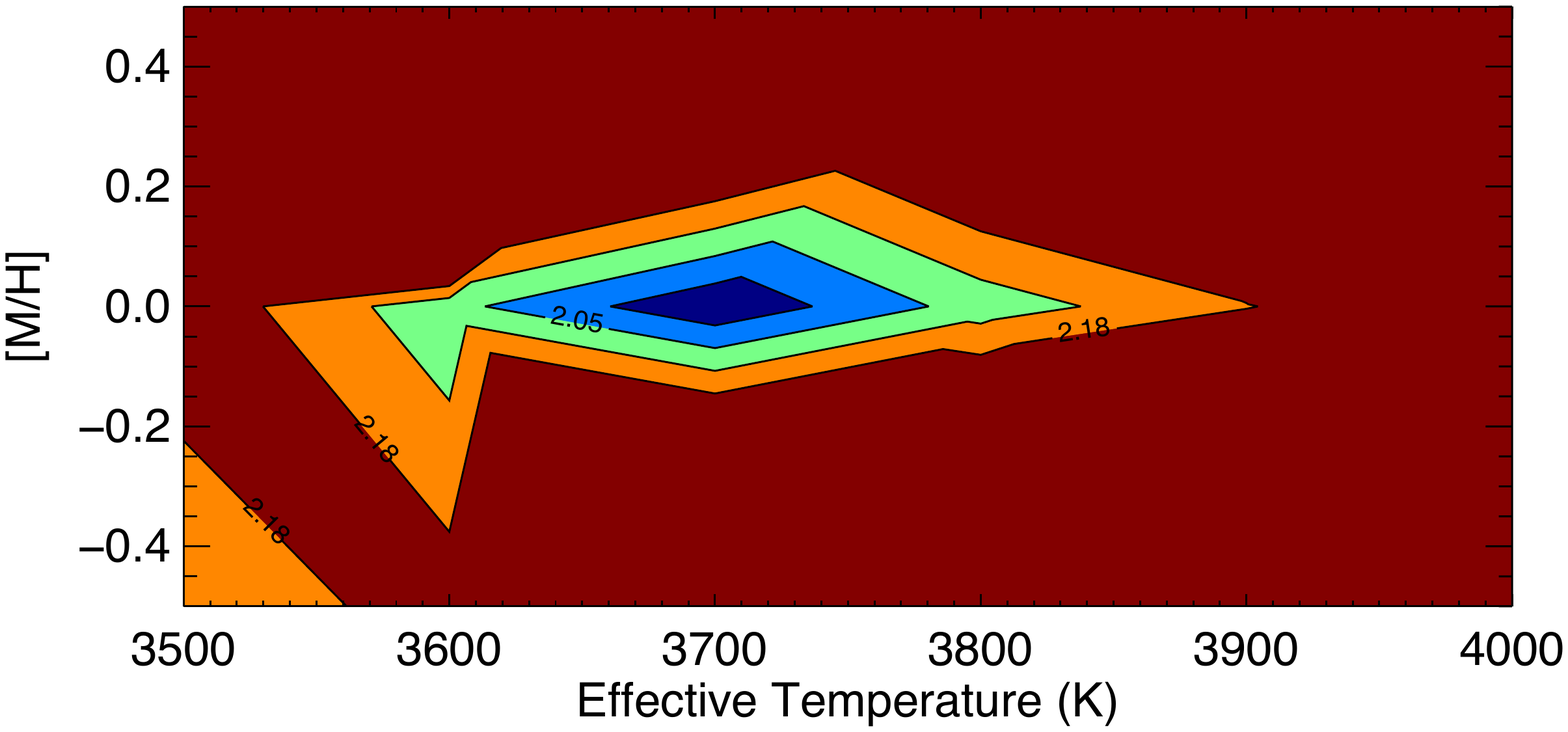}}
\vspace{-1.3cm}
\caption{Example of $\chi^2$ map for different combinations of $\teff$, $\logg$ and [M/H] for GJ 180 (M1.0). The contour plots in the figure shows the calculated $\chi^2$ of the fit based on a grid of$\teff$, $\logg$ and [M/H]. The black lines indicate the $\teff$, $\logg$ and [M/H] with the minimum $\chi^2$.}
\label{Fig3}
\end{figure*}

A second minimisation took place using the values obtained previously as the start point. The new minimisation interpolates in smaller steps (0.1 dex) of $\logg$ and [M/H] but for $\teff$ the step size remains the  same ($\delta$ $\teff$ = 100 K), since it reflects the level of uncertainty in its determination. Decrease in the step size of the interpolation in $\teff$ do not impact the $\chi^2$ minimisation procedure, but
the smaller steps in $\logg$ and [M/H] affect their determination. Figure ~\ref{Fig3} shows contour plots for the lower $\chi^2$ values and represents visually the larger uncertainty regions of the $\logg$ and [M/H] parameters. We retained the stellar parameters (Table 2) corresponding to models for which we obtain the lowest $\chi^2$. Figure~\ref{Fig1} shows such comparison of the observed spectra (black) with that of the best fit model (red) in the optical, whereas,  Fig~\ref{Fig2} shows the similar comparison, but in the J and H band. Unlike, studies by \cite{Passegger2018}, who used $\gamma$-TiO band and Mg I lines, to determine the $\teff$ and metallicity; no weights were applied in our calculation for different parameters during both the steps. For each star a $\chi^2$ map as a function of $\teff$, $\logg$ and [M/H] are obtained. The uncertainty in the parameter space which is 100 K for $\teff$ and 0.3 dex for $\logg$ and [M/H] are calculated by taking the standard deviation of the derived stellar parameters by accepting the 1 $\sigma$ variation from the minimum $\chi^2$. These variations from the minimum $\chi^2$ (shown as different boundaries in Fig~\ref{Fig3}), were calculated using $\chi^2$ statistics.

\begin{figure*}[!htp]
\centering
\subfloat{\includegraphics[height=4.9cm, width=8cm]{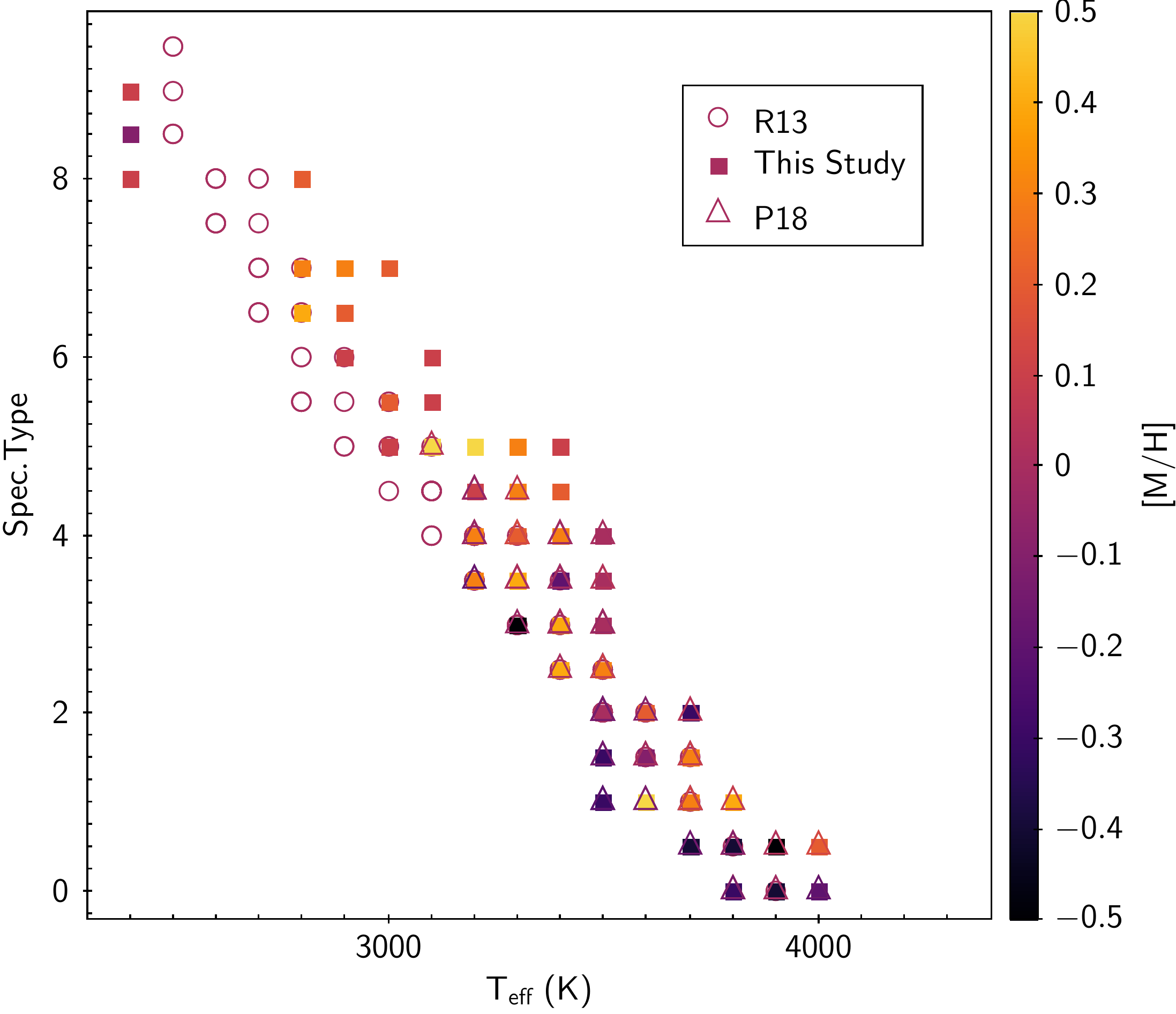}}%
\subfloat{\includegraphics[height=4.9cm, width=8cm]{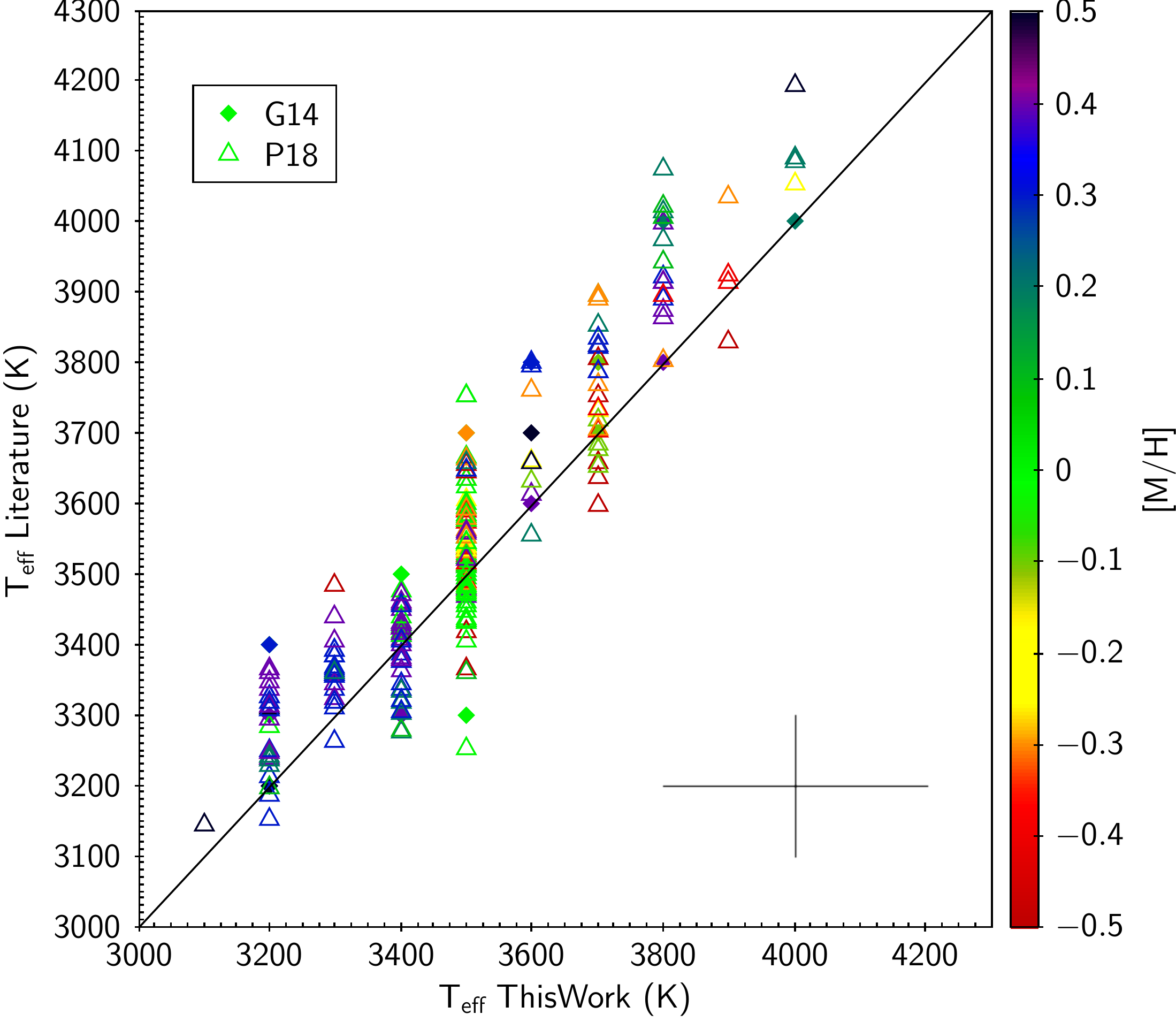}}%
\caption{Right panel: Comparison of $\teff$ versus spectral type relation with our determination with \cite{Passegger2018} (open triangle) and with \cite{Rajpurohit2013} (open circle). Left panel: Comparison of $\teff$ with \cite{Passegger2018}  (open triangle) and with \cite{Gaidos2014} (filled square) as function of metallicity.}
\label{Fig4}
\end{figure*}

\begin{figure*}[!htp]
\centering
\includegraphics[height=16.8cm, width=10.3cm,angle=90]{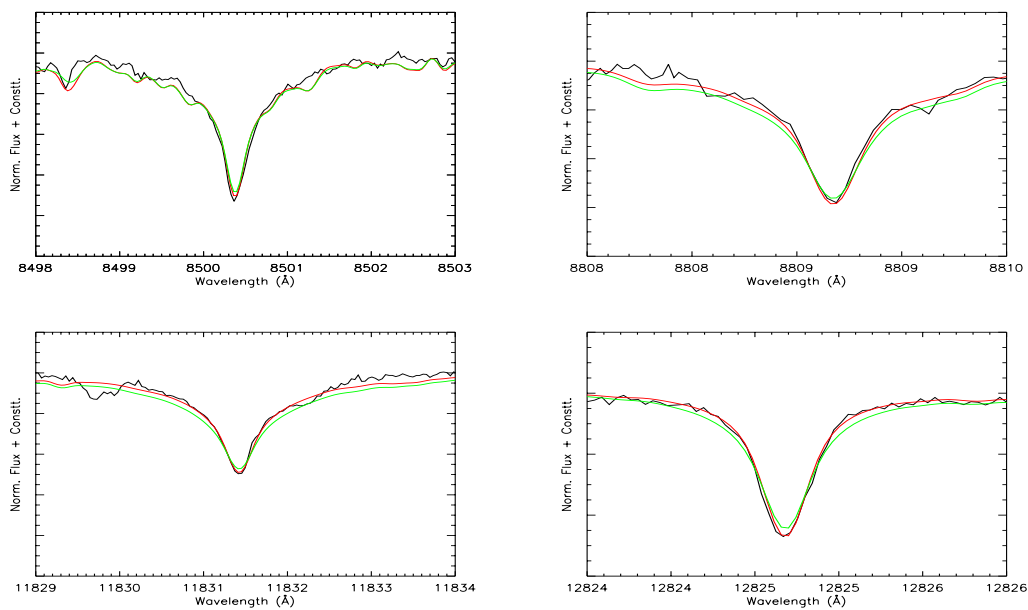}
\vspace{-1.3cm}
\caption{Effect of $\logg$ = 5.0 (red) and $\logg$=5.5 (green) on various atomic lines such as  Ca I (top left and bottom right), Mg I (top right and bottom left) for GJ 180 (M1.0) in black. Their best value of  $\teff$, [M/H] and $\logg$ is given in Table \ref{table1}.  }
\label{Fig5}
\end{figure*}

\section{Analysis and discussion}
\label{analysis}
In order to explore the effect of stellar properties of M dwarfs, we have compared our results with the behavior of $\teff$, as a function of spectral type with \cite{Rajpurohit2013} as shown in Fig~\ref{Fig4} (left panel). \cite{Rajpurohit2013}, claimed that the stars in their samples most probably belong to the thin disk of our Galaxy, and determined the $\teff$ of M dwarfs in their sample by assuming the solar metallicity. \cite{Rajpurohit2013} measured the $\teff$ for the M dwarfs by comparing low resolution visible spectra of M dwarfs for the entire spectral sequence with the Bt-Settl models. In Fig~\ref{Fig4} (left panel) we show the comparison of our results with \cite{Passegger2018} for the common stars in our study. We find that most of the stars in \cite{Passegger2018} sample have the $\teff$ ranging from 3200 K to 4100 K, that is, from early to mid M dwarfs. On the contrary, in our study, we include the M dwarfs cooler than $\teff$ $\leq$ 3000 K where dust and clouds formation are important. We find a systematic difference of about 200 to 300 K between $\teff$ determined by \cite{Passegger2018} and ours for the same spectral type for the common stars. Whereas, our $\teff$ measurements are in good agreement with \cite{Rajpurohit2013} within the 100K. In order to explore further, we compared $\teff$ determination with \cite{Gaidos2014} (hereafter G14) along with \cite{Passegger2018}, for the common stars in our sample as shown in Fig~\ref{Fig4} (right panel). G14 measured the $\teff$ for the M dwarfs by comparing the PHOENIX model spectra with the low resolution visible spectra, as describes by \cite{Mann2013b}. \cite{Passegger2018} compare the observed spectra with the synthetic spectra computed using the PHOENIX-ACES models, which are based on the PHOENIX version by \cite{Husser2013}, and the equation of state by \cite{Barman2001}, as shown in Fig~\ref{Fig4} (right panel). The reason behind the large spread in temperature for each spectral types could be due the fact that \cite{Passegger2018} have used only $\gamma$-TiO band to determine $\teff$, whereas, our $\teff$ come from the overall fit of the observed spectra. Spectra of the PHOENIX-ACE model are computed with the Ames TiO list by \cite{Schwenke1998}, whereas BT-Settl model uses TiO line list by \cite{Plez2008}. We find that TiO bands matched somewhat better with \cite{Plez2008} as compared to what \cite{Passegger2016} have reported. Also, the large differences in $\teff$ could be due to different solar metallicities as mentioned by \cite{Passegger2018} and \cite{Rajpurohit2014}.

\begin{figure*}[!htp]
\centering
\subfloat{\includegraphics[height=4.9cm, width=8cm]{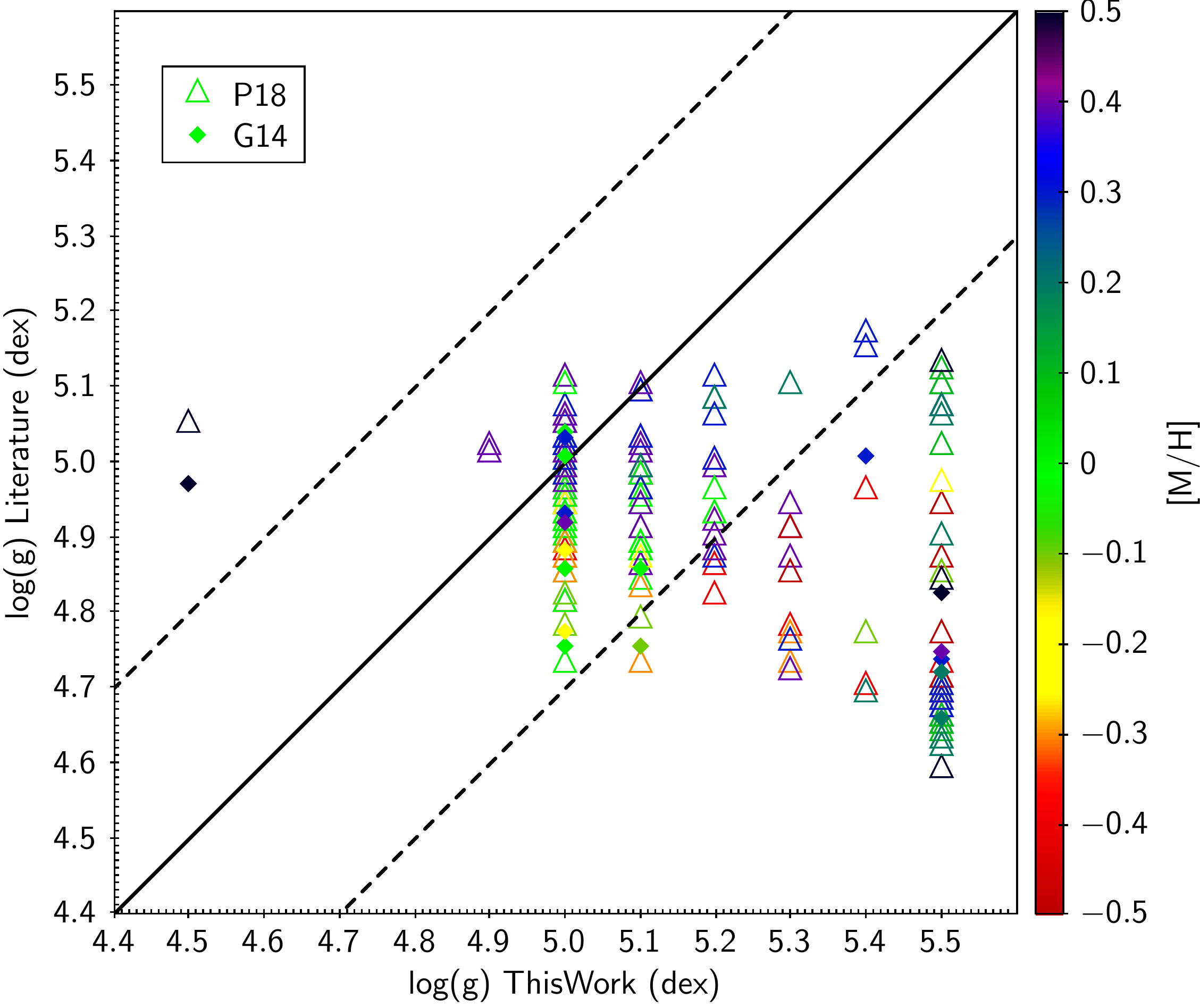}}%
\subfloat{\includegraphics[height=4.9cm, width=8cm]{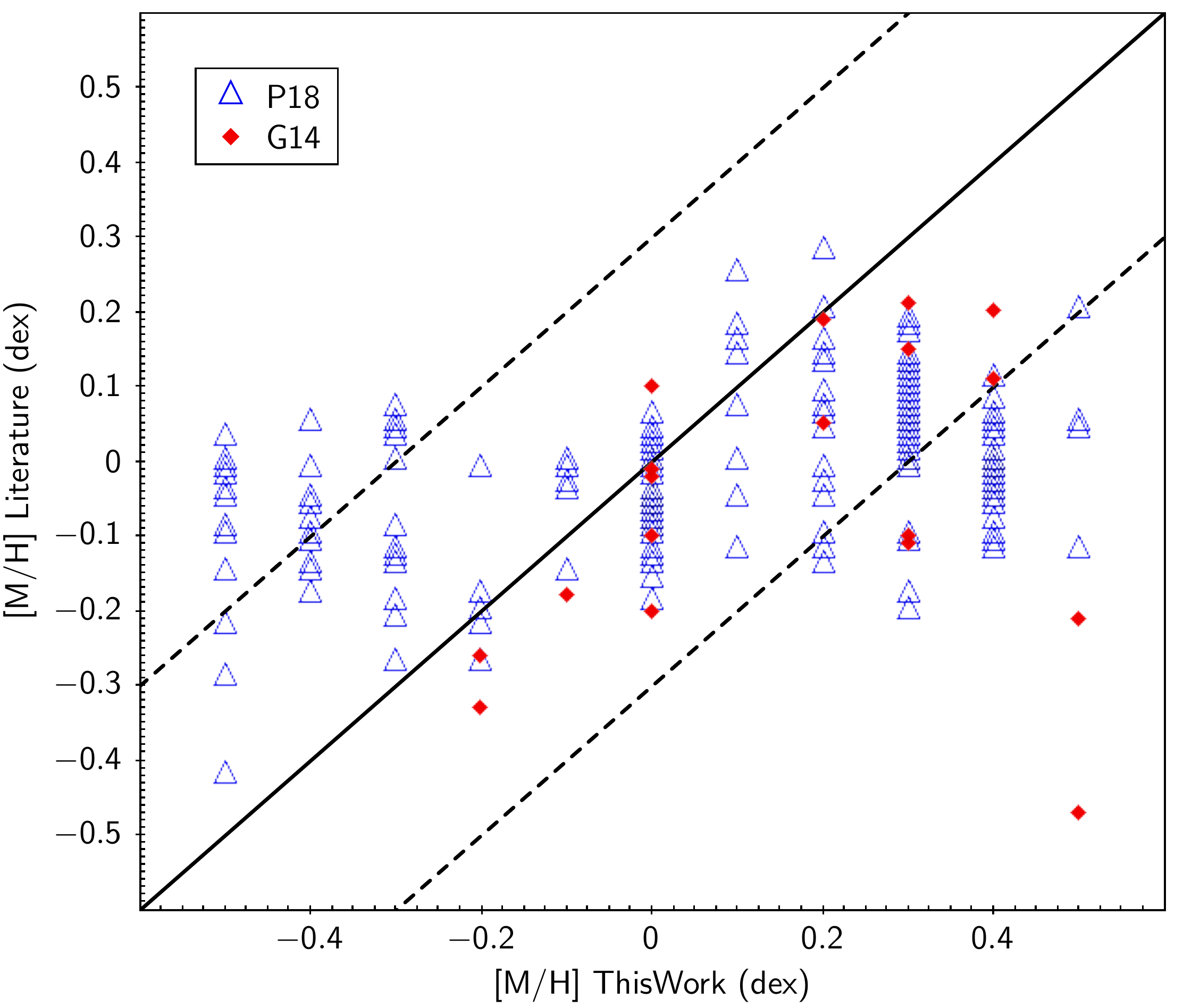}}%
\caption{Left panel: Comparison of $\logg$ from this study with that of \cite{Passegger2018} as function of metallicity. Right panel: Comparison of [M/H] from this study with \cite{Passegger2018} (open trainagle) and with \cite{Gaidos2014} (filled square). The black dotted lines indicates the 1 $\sigma$ deviation in $\logg$ and [M/H].}
\label{Fig6}
\end{figure*}
As suggested by \cite{Gizis1997} and \cite{Casagrande2008}, the typical $\logg$ value of M dwarfs ranges from 4.5 to 5.5. We therefore, explore the effect of gravity using models with $\logg$ from 4.5 to 5.5.  The surface gravity effect in M dwarfs can be seen both in the optical and in the NIR spectra; particularly, in the broadening of atomic lines. Because of the hydrostatic equilibrium in M dwarfs, the photospheric pressure, which is proportional to surface gravity, broadens atomic and molecular features mainly due to H$_2$, He, and H I collisions and due to their higher electron pressure on alkali lines. Various alkali lines, such as, Na I, K I, Ca I, Al I, and Mg I, are stronger at the higher gravity objects. To cross check and confirm the $\logg$ of M dwarfs used in this study, we looked at the gravity-sensitive features both in optical and the NIR spectra (Fig~\ref{Fig5}). 

We checked the overall strength, including the central depth, to confirm $\logg$ obtained using model fittings. These features can be used to discriminate the gravity of M dwarfs to that of sub M dwarfs (sdM) \citep{Gizis1997}. {Most of $\logg$ values ranges from 5.0 to 5.5, which is an expected by interior and evolution models of  \cite{Baraffe2015}. Fig~\ref{Fig6} (left panel) shows the comparison of $\logg$  with  \cite{Passegger2018} indicating difference of around 0.2 to 0.6 dex with our values. The offset in $\logg$ is mainly because \cite{Passegger2018} derived $\logg$ in the same way as \cite{Gaidos2014} by converting the values of $\teff$ into radii, luminosities, and masses using the metallicity-independent empirical relations of \cite{Mann2013b}; which is based on \cite{Baraffe1998} evolutionary models.

Several previous studies have estimated the metallicities of M dwarfs using the wide binary pairs, which have M dwarfs as secondary and the higher mass star as a primary. But in the field M dwarfs, determining global metallicity or individual abundances is a challenge. Most of the attempts to estimate the overall metal content of M dwarfs have been performed at the visible wavelengths. Recent studies by \cite{Rojas2010, Roja2012} and \cite{Newton2015} showed that the mid-resolution NIR spectra can be used to derive the metallicities of M dwarfs, which were based on spectroscopic indices index and equivalent widths of atomic lines. As SED of M dwarfs peaks in the NIR; in this current study, we use both the optical and the NIR high resolution spectra simultaneously to constrain the metallicity. In the observed spectra,  where the molecular absorption is less and atomic features appear clearly, the effect of metallicity can be seen on the strength of various atomic and molecular features. Fig~\ref{Fig6} (right panel) shows the comparison of the [M/H] derived in this study with \cite{Passegger2018} and G14, indicating difference of around 0.2 to 0.4 dex with our values. We find good agreement with the  [M/H] derived by G14, which determine the [M/H] by comparing the BT-Settl model with SNIFS Visible Wavelength Spectra. The synthetic spectrum reproduces fairly well the line profiles of various atomic lines, such as, Ti I, Fe I, Ca I, Mg I, Si I, Mn I, and Al I. Nevertheless, the line strengths of a few atomic lines are not completely reproduced by the models.  An offset of around 0.3 to 0.5 dex in the [M/H] with \cite{Passegger2018} could be due to the systematic errors, which we were unable to eliminate, such as, missing, incomplete or inaccurate opacity sources ( FeH-, OH, and CO-bands), continuum determination and normalisation errors \citep{Neves2014,Rojas2010,Mann2015,Passegger2018}. These sources of errors can result in differences in [M/H] determinations between ours and \cite{Passegger2018} studies.  Moreover, all abundances in the models used by \cite{Passegger2018} are based on the solar composition of \cite{Asplund2009}, instead of the values by \cite{Caffau2011} which is used in the BT-Settl model results in some small differences in the alpha elements \citep{Husser2013,Veyette2016}. These differences in solar abundances become more important in the late-type M dwarfs which is showed by  \cite{Allard2012}. Also, the BT-Settl model includes all the lines in the computation, contributing to differences between the results of the two PHOENIX versions. Thus, a detailed study is required to comment on the impact of varying the alpha parameters on the final spectra and on the metallicity.
\section{Summary}
\label{Summary}
This work showcases the application of the technique developed in \cite{Rajpurohit2012a,Rajpurohit2018} to the CARMENES sample that thoroughly quantifies the uncertainties of the determined parameters. While these values are slightly modified as techniques are refined, they were compared to recent literature and found to be quite similar for the same spectral types. Future work will further explore the impact of finer interpolation steps and their impact on our results. We also studied in detail the properties of 292 early to late M type stars. This is the first time that such a broad wavelength high resolution spectra have been compared to the most updated synthetic spectra computed using the BT-Settl model atmosphere. We have found that there are differences in the stellar parameters of M dwarfs with the others findings based on older models or different opacities from an older setup of Phoenix which are derived either using the optical or the NIR spectra only. In this current study, we have explored the stellar properties of M dwarfs with the high-resolution spectroscopy from optical to NIR simultaneously. We have used  the broad wavelength high resolution spectra (0.52 - 1.71 $\mu$m) to remove any differences and biases, which are from different sets of data as mention by \cite{Bayo2017} and \cite{Rajpurohit2016}. We find that BT-Settl models fit the spectra very well and reproduces the shape of the prominent narrow atomic (K I, Na I, Ca I, Ti I, Fe I, Mg I and Al I) lines, and molecular (TiO, VO, OH, and FeH) features of the objects, where each band is fitted simultaneously. Though there are some discrepancies in reproducing the broadening and the strength of some atomic lines.

We have used the least-squares minimisation technique by comparing observed spectra of M dwarfs in our sample with the BT-Settl models, which give accurate $\teff$ within 100 K and, $\logg$ and [M/H] within 0.3 dex uncertainty. We  provide and compare $\teff$ versus spectral type relation with that of \cite{Passegger2018} and \cite{Rajpurohit2013}. The $\teff$ for the M dwarfs in our samples is extended down to the latest type of M dwarfs, where the dust cloud  begins to form in their atmosphere. The $\teff$ determined using the BT-Settl model agrees well with G14 whereas, it disagree with those of \cite{Passegger2018} with differences up to 300 K for M dwarfs with similar spectral types. In this current study we also showed the comparison of $\logg$ and [M/H] with \cite{Passegger2018} and G14 and found a difference of around 0.2 to 0.8 dex in $\logg$, whereas, for [M/H] it is about 0.5 dex. Thus further exploring the differences in models and parameter space of M dwarfs, our future work will focus on the comparison of synthetic spectra generated using different sets of model atmosphere with both the medium and high resolution spectra of benchmark M dwarfs. 

Comparing the stellar atmospheric parameters obtained by different studies, there are obvious sources which increase the variation between the results. These can take the form of different wavelength ranges or line-lists used for each of the parameters, the used models, the details of the minimisation procedures, the sensitivity of a given parameter to degeneracies in the parameter space, the normalisation procedure, continuum fitting and even the interpolation technique. While some of these have a minimum impact in the final determination (interpolation technique), others can have a significative contribution (models). \cite{Bayo2017} and \cite{Rajpurohit2016} show that the stellar parameters of M dwarfs determined using different sets of data, depend on the various approaches used for that, even for well fitted stars. \cite{Bayo2017} reported that $\teff$, based on the optical data tends to be higher as compared to values computed using the NIR data. Thus, determination of the stellar atmospheric parameters in the mass regime of M dwarfs using the optical and the NIR spectra is a rapidly evolving field, pushed forwards by the recent advances in NIR high-resolution spectrographs.


\begin{acknowledgements}
The research leading to these results has received funding from the French "Programme National de Physique Stellaire" and the Programme National de Planetologie of CNRS (INSU). The computations were performed at the {\sl P\^ole Scientifique de Mod\'elisation Num\'erique} (PSMN) at the {\sl \'Ecole Normale Sup\'erieure} (ENS) in Lyon, and at the {\sl Gesellschaft f{\"u}r Wissenschaftliche Datenverarbeitung G{\"o}ttingen} in collaboration with the Institut f{\"u}r Astrophysik G{\"o}ttingen. G. D. C. Teixeira acknowledges the support by the fellowship PD/BD/113478/2015 funded by FCT (Portugal) and POPH/FSE (EC). This work was supported in part by Funda\c{c}ão para a Ci\^encia e a Tecnologia (FCT) through national funds (UID/FIS/04434/2013) and by FEDER through COMPETE2020 (POCI-01-0145-FEDER-007672). K.R. acknowledges financial support from the ERC Starting Grant "MAGCOW", no. 714196. The result of this research is "based on data from the CARMENES data archive at CAB (INTA-CSIC)". We also want to thank our anonymous referee for useful comments that helped improve the paper.
\end{acknowledgements}
\bibliographystyle{aa}
\bibliography{ref}
\end{document}